\pdfoutput=1
\documentclass[11pt,twoside,a4paper,cmspaper,final,collab]{cms-tdr}
\def\svnVersion{8bc9ddb}\def\svnDate{2022/03/23}\def\cmsCernNoTag{CERN-EP-2021-198}\def\cmsCernDate{\today}\def\cmsMessage{Published in the Journal of High Energy Physics as \href{http://dx.doi.org/10.1007/JHEP03(2022)160}{\doi{10.1007/JHEP03(2022)160}.}}
\begin{document}\cmsNoteHeader{EXO-20-003}

\DeclareRobustCommand{\PS}{{\HepParticle{S}{}{}}\xspace}
\newcommand{\HToSS}{\ensuremath{\PH\to\PS\PS}\xspace}
\newcommand{\NT}{\ensuremath{N_{\mathrm{j}}^{\text{dis}}}\xspace}
\newcommand{\IP}{\ensuremath{\hat{\mathrm{IP}}^\mathrm{2D}_\text{sig}}\xspace}
\newcommand{\TA}{\ensuremath{\hat{\Theta}^\mathrm{2D}}\xspace}
\newcommand{\AL}{\ensuremath{\alpha_\text{max}}\xspace}
\newcommand{\cbf}{\ensuremath{\mathcal{B}}\xspace}
\newcommand{\To}{\scalebox{1.1}{\ensuremath{\mathcal{\to}}}}
\newcommand{\dxy}{\ensuremath{d_{xy}}\xspace}
\newcommand{\sdxy}{\ensuremath{\sigma_{\dxy}}\xspace}
\newcommand{\Ri}{\ensuremath{R_i}\xspace}
\newcommand{\RiTZ}{\ensuremath{\Ri^{\PQt\to\PZ}}\xspace}
\newcommand{\sirel}{\ensuremath{\sigma^{i}_\text{rel}}\xspace}

\cmsNoteHeader{EXO-20-003}

\title{Search for long-lived particles produced in association with a \texorpdfstring{$\PZ$}{Z} boson in proton-proton collisions at \texorpdfstring{$\sqrt{s} = 13\TeV$}{sqrt(s) = 13 TeV}}

\date{\today}

\abstract{
A search for long-lived particles (LLPs) produced in association with a \PZ boson is presented. The study is performed using data from proton-proton collisions with a center-of-mass energy of 13\TeV recorded by the CMS experiment during 2016--2018, corresponding to an integrated luminosity of 117\fbinv. The LLPs are assumed to decay to a pair of standard model quarks that are identified as displaced jets within the CMS tracker system. Triggers and selections based on \PZ boson decays to electron or muon pairs improve the sensitivity to light LLPs (down to 15\GeV{}). This search provides sensitivity to beyond the standard model scenarios which predict LLPs produced in association with a \PZ boson. In particular, the results are interpreted in the context of exotic decays of the Higgs boson to a pair of scalar LLPs ($\HToSS$). The Higgs boson decay branching fraction is constrained to values less than 6\% for proper decay lengths of 10--100\mm and for LLP masses between 40 and 55\GeV. In the case of low-mass ($\approx$15\GeV{}) scalar particles that subsequently decay to a pair of \PQb quarks, the search is sensitive to branching fractions $\cbf(\HToSS) < 20\%$ for proper decay lengths of 10--50\mm. The use of associated production with a \PZ boson increases the sensitivity to low-mass LLPs of this analysis with respect to gluon fusion searches. In the case of 15\GeV scalar LLPs, the improvement corresponds to a factor of 2 at a proper decay length of 30\mm.}

\hypersetup{
pdfauthor={CMS Collaboration},
pdftitle={Search for long-lived particles produced in association with the Z boson in proton-proton collisions at sqrt(s)= 13 TeV},
pdfsubject={CMS},
pdfkeywords={CMS, long-lived particle, LLP, low mass, Higgs boson, Z boson, lepton, bottom quark}}

\maketitle

\section{Introduction}

Long-lived particles (LLPs) with macroscopic decay lengths arise in many extensions of the standard model (SM) of elementary particles.
The long list of classes of models where LLPs are found includes supersymmetry~\cite{Arvanitaki2013,ArkaniHamed:2012gw,GIUDICE1999419,BARBIER20051,Csaki:2013jza,Fan:2011yu,Fan:2012jf,Burdman:2006tz}, little Higgs~\cite{Cai:2008au}, twin Higgs~\cite{Chacko:2005pe}, hidden valley models~\cite{Strassler:2006im,Strassler:2006ri}, and dark sector models~\cite{Baumgart:2009tn,Kaplan:2009ag,Dienes:2011ja,Dienes:2012yz}.
Long-lived particles may also play a role in explaining baryogenesis~\cite{Cui:2014twa} or in accommodating neutrino masses~\cite{Batell:2016zod}.
The LLPs produced in proton-proton ($\Pp\Pp$) collisions at the CERN LHC could decay into SM particles away from the interaction point of the colliding beams.

Models which propose a form of neutral naturalness~\cite{Chacko:2015fbc} are of particular interest.
In these models, the hierarchy problem is resolved through the existence of a partner of the top quark that is not color-charged under the SM $\mathrm{SU}(3)_\text{c}$ group but is instead charged under a mirror color group of a hidden sector, and that thereby evades the stringent experimental bounds on strongly produced particles.
The Higgs boson (\PH) may then decay into long-lived bound states of the mirror color group, which subsequently decay into SM particles through the kinetic mixing with the Higgs boson itself.
The LLP decays through an off-shell Higgs boson predominantly to a pair of quarks~\cite{Curtin:2013fra}.
Such decays will therefore manifest themselves as displaced jets that originate from the hadronization of the final-state quarks.

\begin{figure}[htb]
      \centering
      \includegraphics[width=0.48\linewidth]{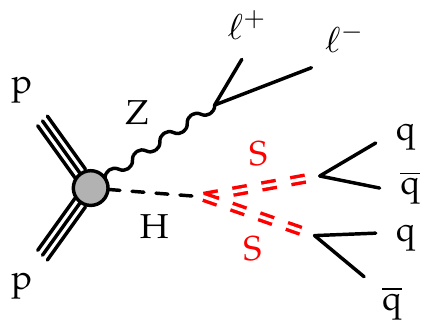}
      \caption{Feynman diagram of a simplified model for the Higgs boson decaying to a pair of long-lived scalar particles (\PS).
      The Higgs boson is produced in association with a \PZ boson, where the \PZ boson decays to a pair of leptons.
      The long-lived scalars decay to a pair of quark jets (\PQq).}\label{fig:twinhiggs}
  \end{figure}

Previous searches at the LHC for displaced jets have typically relied on jets with large transverse momentum (\pt), whose presence is required in order to pass the trigger requirements.
Those searches thereby have had limited sensitivity to decays to displaced jets with low \pt.
Here we present an alternative approach that exploits a new LLP search channel, namely the associated production of LLPs with a \PZ boson, where prompt leptons (electrons and muons) provide an effective trigger for events with low-\pt jets.
Figure~\ref{fig:twinhiggs} shows a simplified model where the Higgs boson is produced in association with a \PZ boson, and subsequently decays to a pair of long-lived scalar particles (\PS).
To tag a jet from an \PS decay as being displaced, we rely on information from the CMS tracking system, which provides the greatest sensitivity for mean proper decay lengths of $\approx$10\unit{cm}.
Searches for LLPs produced in Higgs boson decays have been recently performed at $\sqrt{s} = 13\TeV$ by the CMS~\cite{CMS:2020iwv,CMS:2021juv} and ATLAS experiments~\cite{ATLASMS,ATLASCALO,ATLASTRACKERMS,ATLAS:2021jig}, where they have focused on the production of the Higgs boson through gluon fusion.

The paper is organized as follows.
A brief description of the CMS detector is given in Section~\ref{sec:detector}.  Section~\ref{sec:samples} describes the data and the simulated events used.
Section~\ref{sec:eventsel} explains the displaced-jet identification strategy and the event selection.
The estimation of the background is described in Section~\ref{sec:background}, and the
treatment of systematic uncertainties is given in Section~\ref{sec:systematics}.
Results of the search are described in Section~\ref{sec:results}, which is followed by a summary of the paper in Section~\ref{sec:summary}.

Tabulated results are provided in the HEPData record for this analysis~\cite{hepdata}.

\section{The CMS detector}\label{sec:detector}

The central feature of the CMS apparatus is a superconducting solenoid of 6\unit{m} internal diameter, providing a magnetic field of 3.8\unit{T}.
Within the solenoid volume are a silicon pixel and strip tracker, a lead tungstate crystal electromagnetic calorimeter (ECAL), and a brass and scintillator hadron calorimeter (HCAL), each composed of a barrel and two endcap sections.
Forward calorimeters extend the pseudorapidity ($\eta$) coverage provided by the barrel and endcap detectors.
Muons are measured in gas-ionization detectors embedded in the steel flux-return yoke outside the solenoid.

The silicon tracker measured charged particles within the pseudorapidity range $\abs{\eta} < 2.5$ ($\abs{\eta} < 3.0$) during the LHC running period in 2016 (2017--2018).
For the data used in this paper, the silicon tracker consisted of 1440 (1856) silicon pixel detector modules during the 2016 (2017--2018) running period and 15,148 silicon strip modules throughout the 2016--2018 data-taking period.
For jet constituents with $1 < \pt < 10\GeV$ and $\abs{\eta} < 1.4$ ($\abs{\eta} < 3.0$), the track resolutions are typically 1.5\% in \pt and 25--90\mum (20--75\mum) in the transverse impact parameter during 2016 (2017--2018)~\cite{Chatrchyan:2014fea}.

In the region $\abs{\eta} < 1.74$, the HCAL cells have widths of 0.087 in $\eta$ and 0.087 in azimuth ($\phi$).
In the $\eta$-$\phi$ plane, and for $\abs{\eta} < 1.48$, the HCAL cells map onto $5{\times}5$ arrays of ECAL crystals to form calorimeter towers projecting radially outwards from a point close to the nominal interaction point.
For $\abs{\eta} > 1.74$, the coverage of the towers increases progressively to a maximum of 0.174 in $\Delta\eta$ and $\Delta\phi$.
Within each tower, the energy deposits in ECAL and HCAL cells are summed to define the calorimeter tower energies, which are subsequently used to provide the energies and directions of hadronic jets.

Events of interest are selected using a two-tiered trigger system~\cite{Khachatryan:2016bia}.
The first level, composed of custom hardware processors, uses information from the calorimeters and muon detectors to select events at a rate of around 100\unit{kHz} within a time interval of less than 4\mus~\cite{Sirunyan:2020zal}.
The second level, known as the high-level trigger, consists of a farm of processors running a version of the full event reconstruction software optimized for fast processing, and reduces the event rate to around 1\unit{kHz} before data storage.
A more detailed description of the CMS detector, together with a definition of the coordinate system used and the relevant kinematic variables, can be found in Ref.~\cite{Chatrchyan:2008zzk}.

\section{Data and simulated samples}\label{sec:samples}
This search uses a sample of $\Pp\Pp$ collisions collected in 2016--2018 at $\sqrt{s} = 13\TeV$, corresponding to an integrated luminosity of 137.6\fbinv.
During part of the 2016 data-taking period there was a decreased tracking efficiency within the tracker system that resulted in spurious displaced jets; therefore collision events occurring during the first 20\fbinv, when the effect took place, are excluded from this search.
After removing the affected events, the effective total integrated luminosity used by the search is 117\fbinv.

Events in this search were recorded using lepton-based triggers.
One set of triggers requires an isolated pair of either electrons or of muons~\cite{Sirunyan_mu_2021}.
Another set of triggers requires electron-muon pairs, where both the electron and muon are required to be isolated.
The electron-muon pair triggers, which are included as part of the study of the background, are discussed later in this paper.

The performance of the analysis of $\PZ\PH$ events containing LLPs is evaluated using simulated events.
Quark- and gluon-initiated associated Higgs boson production processes are generated with \POWHEG~2.0~\cite{Nason:2004rx,Frixione:2007vw,Alioli:2010xd,Alioli:2009je,Re:2010bp,Luisoni:2013cuh} at leading order (LO) and next-to-LO (NLO) precision, respectively. 
The total $\PZ\PH$ production cross section is 0.88\unit{pb}, taken from Ref.~\cite{deFlorian:2016spz}.
Higgs boson decays to long-lived scalars are generated with \PYTHIA~8.226~\cite{Sjostrand:2014zea}.
The LLP is simulated as a generic scalar particle with a 100\% branching fraction to
either heavy quarks (studied here using \PQb quarks) or light quarks (studied here using \PQd quarks).
The mass of the scalar particle $m_\PS$ corresponds to 15, 40, or 55\GeV.
The proper mean lifetime $c\tau_\PS$ is varied within the range 1--1000\mm, and evaluated at 13 different points across this range.
These ranges are consistent with the recommendations of the LHC Higgs Cross Section Working Group~\cite{deFlorian:2016spz}.

The Drell--Yan process (DY), which is the main background in this search, is simulated at NLO in quantum chromodynamics (QCD) using the \MGvATNLO~2.4.2~\cite{MADGRAPH} generator with up to two additional partons in the final state at the matrix element (ME) level.
The corresponding cross section is calculated with \FEWZ~v3.1b2~\cite{PhysRevD.86.094034} at next-to-NLO (NNLO) in QCD and NLO precision in electroweak theory.
The top quark-antiquark (\ttbar) background is simulated with NLO precision in QCD using the \MGvATNLO generator, and its cross section is obtained from the \textsc{TOP++}~v2.0~\cite{Czakon:2011xx} prediction that includes NNLO corrections in QCD and resummation of the next-to-next-to-leading-logarithm soft-gluon terms.
The single top quark processes are simulated at NLO in QCD via either \sloppy \POWHEG~2.0 or \MGvATNLO and their cross sections are computed, at the same order of precision, using \textsc{hathor}~v2.1~\cite{Kant:2014oha}.
For the simulated samples corresponding to the 2016 (2017--2018) data-taking periods, the NNPDF~v3.0 (v3.1) NLO (NNLO) parton distribution functions, PDFs, are used~\cite{Ball:2014uwa,Ball:2017nwa}.
The CUETP8M1 tune~\cite{Khachatryan:2015pea} is used for the simulated samples corresponding to the 2016 data-taking period, while the CP5 tune~\cite{Sirunyan:2019dfx} is used for the 2017 and 2018 simulated data.
For processes generated at NLO (LO) in QCD with the \MGvATNLO generator, events from the ME characterized by different parton multiplicities are merged via the FxFx~\cite{Alwall:2007fs} (MLM~\cite{Frederix:2012ps}) prescription.
The simulated events at the ME level for both the signal and background processes are interfaced with \PYTHIA~8.226 or a later version to simulate the shower and hadronization of partons in the initial and final states, along with the underlying event description.

For all simulated processes, the detector response is simulated using a detailed description of the CMS detector based on \GEANTfour~\cite{GEANT4}.
Object and event reconstruction are performed with the same algorithms as are used for the data.
Minimum bias events are superimposed on each simulated hard scattering event to reproduce the effect of extra $\Pp\Pp$ interactions within the same or neighboring bunch crossing as the recorded event (pileup).
The frequency distribution of the additional events is adjusted to match that observed during each data-taking period.

\section{Search strategy and selections}\label{sec:eventsel}
The basic strategy of this search is to use the displaced jet multiplicity (\NT) in the event to distinguish the signal from the background processes, where a displaced jet is defined as a jet that passes specified selections made on the three tagging variables described later in this section.
Signal events typically contain $\NT \geq 2$, while SM background processes exhibit a sharply falling distribution in \NT.

Backgrounds include events with displaced jets arising from decays of long-lived SM particles, nuclear interactions with the tracker material, photon conversions, and mismeasurement of jet constituents.
Since the simulation may not capture these effects perfectly, we use a strategy based on control samples in data to estimate the number of misidentified displaced jets from SM background processes, by considering control samples of low-\pt opposite-sign dilepton pairs (modeling the dominant SM \PZ boson production) and of different-flavor ($\Pe\Pgm$) pairs (modeling the subdominant contribution from \ttbar and single top quark production).
Rare background processes, including SM multiboson production, are estimated adequately from simulation, as their contribution to the signal sample is small.

Events are required to have a primary interaction vertex (PV), corresponding to the hardest scattering in the event, evaluated using tracking information alone, as described in Section 9.4.1 of Ref.~\cite{CMS-TDR-15-02}.

The energy of each electron is determined from a combination of the electron momentum at the primary interaction vertex, to be defined below, as determined by the tracker, the energy of the corresponding ECAL cluster, and the energy sum of all bremsstrahlung photons spatially compatible with originating from the electron track~\cite{Sirunyan_2021}.
The momentum of each muon is obtained from the curvature of the corresponding track~\cite{Sirunyan:2018mu}.
Electron pairs are required to have a minimum \pt of 25 (15)\GeV for the leading (subleading) candidate, while for muon pairs the thresholds are 25 (12)\GeV.
Electrons and muons are required to have $\abs{\eta} < 2.5$ and $\abs{\eta} < 2.4$, respectively.
The resulting efficiencies for electrons and muons, measured in SM \PZ boson simulation and data, are above 90\%.

Jets are reconstructed offline from the energy deposits in the calorimeter towers, clustered using the anti-\kt algorithm~\cite{Cacciari:2008gp, Cacciari:2011ma} with a distance parameter of 0.4.
In this process, the contribution from each calorimeter tower is assigned a momentum, the absolute value and the direction of which are given by the energy measured in the tower and the coordinates of the tower relative to those of the PV.
The raw jet energy is obtained from the sum of the tower energies, and the raw jet momentum by the vectorial sum of the tower momenta, which results in a nonzero jet mass.
The raw jet energies are then corrected to establish a uniform relative response of the calorimeter in $\eta$, and a calibrated absolute response in \pt.
Jets are required to have a minimum \pt of 35\GeV and to fall within the silicon strip tracker acceptance $\abs{\eta} < 2.4$.
Cuts are applied to suppress contributions from electronic noise, electrons, and muons.
`High purity' tracks with $\pt \geq 1\GeV$~\cite{Chatrchyan:2014fea} are subsequently matched in ($\eta$, $\phi$) to a given jet if the distance between the track and the jet axis $\Delta R = \sqrt{\smash[b]{(\Delta\eta)\,^2 + (\Delta\phi)\,^2}} < 0.4$.
The direction of each track included in the matching is extrapolated to the inner surface of the calorimeter.
The matched tracks are used to evaluate whether the jet is displaced.

Additionally, there must be at least one opposite-sign, same-flavor lepton pair in the event.
The lepton pair is required to have an invariant mass between 70 and 110\GeV and a \pt of at least 100\GeV.
The selection requirement on the \pt of the lepton pair serves to suppress DY events with a typically low dilepton \pt, and to enhance the relative contribution of associated Higgs boson production.
We require that there be no additional leptons with \pt larger than 15\GeV, and at least one jet, which defines the preselection sample.

In addition to the search sample, we define two control samples that are used for the background estimation presented in Section~\ref{sec:background}.
The \PZ boson control sample has the same requirements as the signal sample, except the dilepton \pt must range between 10 and 100\GeV.
A second control sample is used to estimate the background from \ttbar and single top quark production, and is hereafter referred to as the top quark control sample.
The top quark control sample is selected by requiring different-flavor lepton pairs with a \pt of at least 10\GeV.

\begin{figure}[ht]
    \centering
    \includegraphics[width=0.32\linewidth]{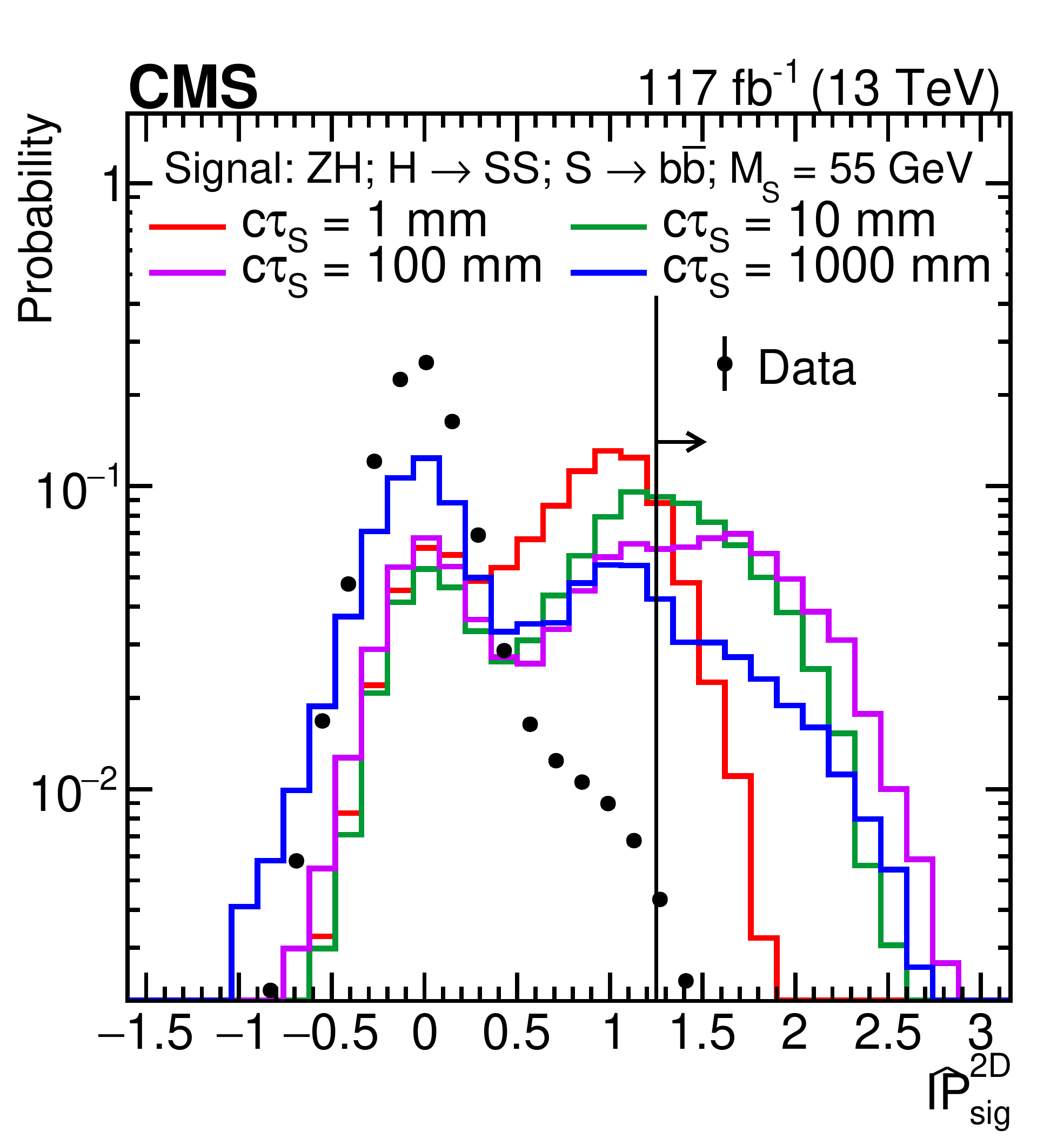}
    \includegraphics[width=0.32\linewidth]{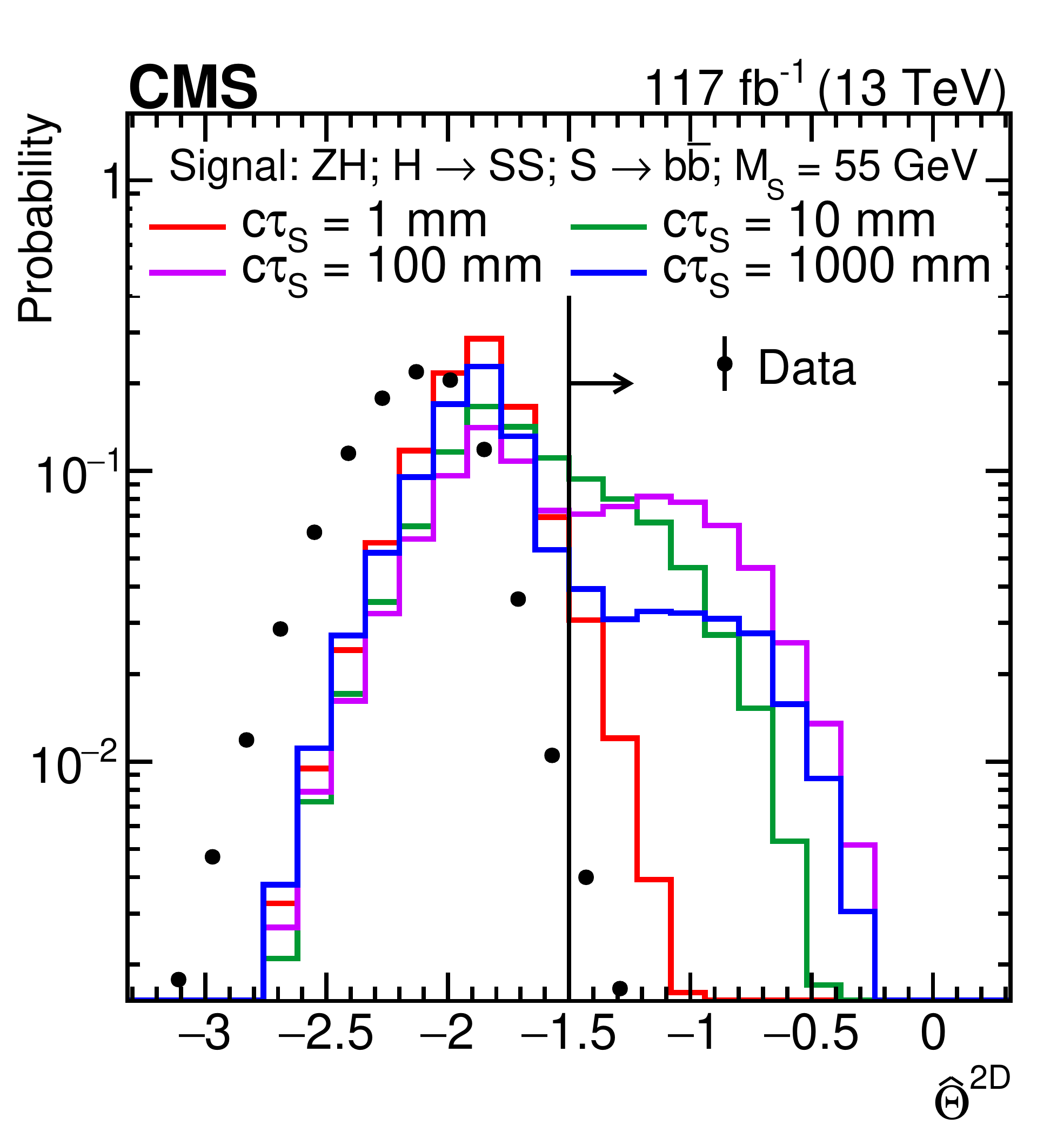}
    \includegraphics[width=0.32\linewidth]{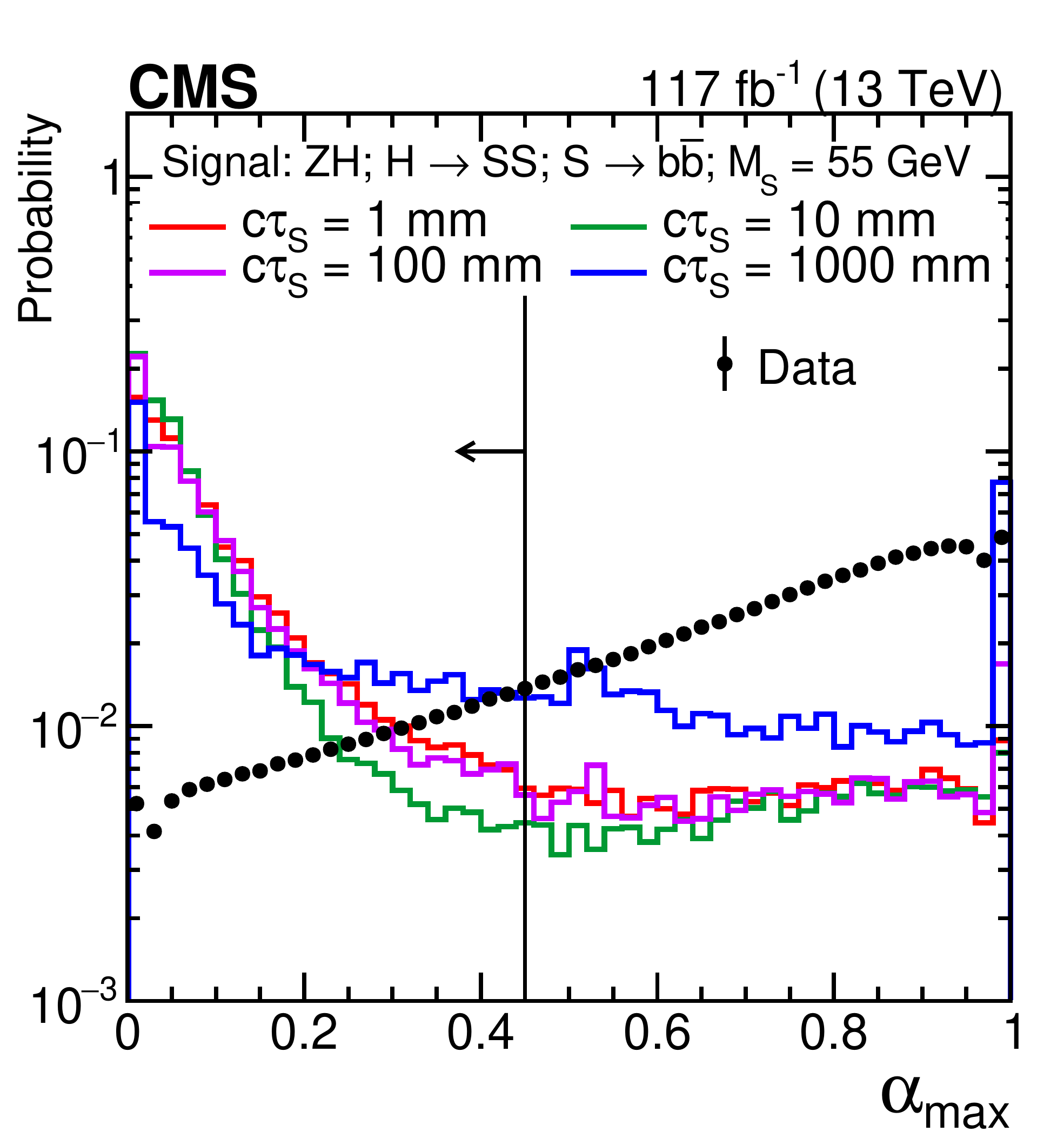}
    \caption{Distributions of the three tagging variables for data and for four signal samples,
    where the decay lengths of the signal range from 1 to 1000\mm.
    The left figure displays the distributions for the \IP tagging variable, while the center and right figures display the distributions for the \TA and \AL tagging variables, respectively.
    Overlaid on the figures is a line with an arrow pointing to the region of where values of the variable are used to aid in distinguishing possible displaced jets from background jets.}\label{fig:TVPlots}
\end{figure}

In order to identify displaced jets, the properties of the tracks associated with jets are used to
calculate three displacement variables known as ``tagging variables,'' as previously reported by the CMS Collaboration~\cite{displacedjets1}.
Distributions of the tagging variables in data and simulation are shown in Fig.~\ref{fig:TVPlots}.
The first tagging variable is the jet impact parameter significance (\IP), defined as the median of the logarithm (base 10) of the impact parameter significance ($\dxy/\sdxy$) of all the tracks matched to the jet, with \dxy being the track transverse impact parameter and \sdxy its uncertainty.
The SM background processes, largely composed of prompt jets whose tracks have \dxy values of the order of \sdxy, exhibit an \IP value peaked near zero.
On the other hand, displaced decays of LLPs tend to have jets with larger \IP values.
For $\IP>1.25$, the chosen threshold, it means that the transverse impact parameter is about $18\sdxy$ and thus highly significant.
The second tagging variable is the jet transverse angle (\TA), which is calculated as the median logarithm (base 10) of the angle between the track direction and the vector connecting its innermost hit in the silicon tracker to the PV.
Prompt jets have \TA near zero as the vector connecting the PV to the track's innermost hit tends to be aligned with the jet-axis.
In the case of LLPs, the directions of these vectors are not necessarily aligned, thus corresponding to larger angles.
The last tagging variable is \AL, which is calculated as follows: for each PV candidate ($v_i$), we define $\alpha$ as
 \begin{equation}
 \label{eq:AM}
   \alpha(v_i) = \frac{\sum_{\text{tracks} \in v_i}  \pt^\text{track}}{\sum_{\text{all tracks}} \pt^\text{track}},
 \end{equation}
the ratio of the summed-\pt for the tracks within the jet that are associated with that particular vertex to the total summed-\pt for all tracks within the jet.
The \AL variable corresponds to the maximum $\alpha$ value across all reconstructed vertices.
The LLPs form displaced jets that typically exhibit \AL values near zero, as the tracks belonging to these jets do not originate from any collision vertex.

Finally, we determine requirements for each of the three tagging variables that together maximize the discovery reach for this search.
Jets are identified as displaced when $\IP>1.25$, $\TA>-1.5$, and $\AL<0.45$.
These selections were determined via a multi-step optimization, where the selection cuts correspond to a relatively large signal significance, as calculated using the
Punzi significance method~\cite{Punzi}.
The statistical uncertainty on the background was considered during the optimization.
The number of displaced jets $\NT$ is then used to distinguish the signal from background, where an excess of events where $\NT \geq 2$ would indicate the presence of a signal.
The signal efficiency with respect to the baseline requirement to select a \PZ boson in association with a jet was measured.
The resulting efficiency for signal events in the $\NT \geq 2$ bin, was found to be 5.4, 16.1, and 19.5\% (5.9, 19.8, and 22.3\%) in the 4\PQb (4\PQd) quarks final state for masses of 15, 40, and 55\GeV, respectively, and for a mean proper decay length of 10\mm.
The corresponding background rejection, inferred from simulation, was found to be $7.6 \times 10^5$.

\section{Background estimation}\label{sec:background}

In the case of the two main sources of background: DY and top quark (including \ttbar and single top) production, the background estimation is carried out using the control samples defined in the previous section.
We estimate the contribution of these two sources of background to the search sample by performing a maximum likelihood fit binned in \NT using the control and signal samples simultaneously.
The ratio of the background yield in the search sample to the control sample for each \NT bin $i$ is defined as the transfer factor, and is constrained to its value measured from the simulation samples.
Three \NT bins are considered: 0, 1, and 2 or more.
The DY transfer factors ($\Ri^\PZ$) are defined as the ratio of the yields in the $i$th bin in the search sample to the corresponding bin in the \PZ boson control sample, where both yields are obtained from the DY simulation.
Similarly, the transfer factors for the top quark background processes ($\Ri^\PQt$) are the ratios of the yield in the $i$th bin in the search sample to the corresponding bin in the top quark control sample, where both yields are obtained from the simulation of \ttbar and single top quark processes.
A third set of transfer factors (\RiTZ) is used to account for the contamination from top quark background processes in the \PZ boson control sample.
The \RiTZ variable is defined as the ratio of the yields in the $i$th bin in the \PZ boson control sample to the corresponding bin in the top quark control sample, where both yields are obtained from the simulation of \ttbar and single top quark processes.

The $\Ri$ values, their associated uncertainties, as well as the yields in the control and signal samples are used as the ingredients in a likelihood model, which fits the background contributions and the signal yield.
The fit is performed simultaneously across the signal and control samples.
The background yields in the search sample, for each displaced jet multiplicity bin $i$, are the sum of the yields from the individual background processes.
The contributions from the main background processes (DY and top quark production) in the signal sample are estimated as the product of the $\Ri$ values and a set of freely floating parameters, which represent the yield in each of the $i$ bins for a specific background process in the corresponding control sample.
By performing a fit in the signal and control samples we are able to obtain simultaneously for each process the yields in the control samples and the values of the transfer factors, and from these the estimated total background yield in the signal sample.

The systematic uncertainties affecting the $\Ri$ transfer factors, which are described in Section~\ref{sec:systematics}, are included in the simultaneous fit.
Each systematic uncertainty source is modeled as a nuisance parameter with a log-normal constraint in the likelihood.
The corresponding statistical uncertainty in $\Ri$ is propagated and treated as uncorrelated among the bins.

For rare background processes such as SM multiboson events or SM Higgs boson decays, the contributions to the search sample are taken directly from the simulation, and correspond to less than 2\% of the total background yield.

\section{Systematic uncertainties}\label{sec:systematics}

The main sources of systematic uncertainty are related to the possibility that the transfer factors obtained from the simulation do not accurately reflect the data, and the values are detailed in Table~\ref{tab:SignalSystematics}.

In order to check the validity of the treatment of systematic uncertainties and rule out other possible systematic effects, we construct seven independent validation samples (VS, Table~\ref{tab:VRs}), by inverting the requirement on one or more of the displaced jet tagging variables, where the signal contamination in each validation sample is negligible.
We perform the background estimation procedure for each validation sample in the same manner as in the main search sample, as detailed in Section~\ref{sec:background}.
If the background estimation is working as intended and the treatment of the systematic uncertainties is adequate, we should find that the binned-likelihood model will accurately reproduce the displaced jet multiplicity distribution in all the validation samples.
No significant deviations are observed when performing the fit in the validation samples (see Fig.~\ref{fig:jetmult}), thus validating the background estimation method and the treatment of the systematic uncertainties.

\begin{table*}[!tbph]
    \centering
    \topcaption{The systematic uncertainties in the background and signal
        yield expectations.
        Dashes indicate that the systematic effect is not applicable or is negligible.}
    \label{tab:SignalSystematics}
    \begin{tabular}{lcc}
        \hline
        Uncertainty source & Signal (\%) & Background (\%) \\
        \hline
        Luminosity                                 & 1.6    & ${<}0.01$ \\
        Dilepton trigger scale factors             & 1      & \NA \\
        Lepton energy scale                        & 0.5--1.0 & \NA  \\
        Lepton ID and ISO scale factors            & 1--2   & \NA       \\
        Jet energy scale                           & 4--8   & 2--4  \\
        Tagging variable shape correction          & 1--20  & 1--5  \\
        Transfer factor (simulated background sample size)  & \NA    & 1--90 \\
        Simulated signal sample size & 5--30       & \NA       \\
        Control sample (statistical)               & \NA    & 0.1--5.0\\
        \hline
    \end{tabular}
\end{table*}

The simulated expected yields for signal and the small contributions from rare background processes are directly affected by the uncertainty in the integrated luminosity, estimated to be 1.6\%~\cite{lumiUnc16,lumiUnc17,lumiUnc18}.
Lepton energy scale uncertainties affect the efficiency of the \PZ boson mass constraint in the search sample; these are varied according to values extracted from a study of leptonic \PZ boson decays and result in 0.5--1.0\% uncertainties in the signal yields.
Lepton efficiency uncertainties are extracted from a tag-and-probe analysis~\cite{CMS:2010svw} and are 1--2\%.
Uncertainties in the jet energy scale affect the jet \pt selection efficiency at the level of 4--8\% (2--4\%) for signal (background).
Varying the shape of the displaced jet tagging variable distributions to account for the measured simulation mismodeling results in an uncertainty of 1--20\% (1--5\%) for signal (background), depending on the multiplicity bin.
The statistical uncertainty in the transfer factors increases with the displaced jet multiplicity and reaches a maximum of 90\% in the $\NT\geq 2$ bin, thus representing the largest source of uncertainty in the search.
The statistical uncertainty contribution in the signal sample from the control samples is due to the finite size of the control samples in data and is found to be 0.1--5\%.
The statistical uncertainty in the yields of the background processes taken from simulation is found to be negligible.
The signal simulation statistical uncertainty is found to be 1--10\% in the signal sample, depending on the signal model and the \NT bin.

\begin{table*}[h!]
\centering
  \topcaption{Summary of the track-based displaced-jet tagging requirements to define the seven validation samples, VS$_{1}$ through VS$_{7}$, and the signal sample Sig S.}
    \label{tab:VRs}
    \begin{tabular}{cccc}\hline
        & \IP$_\text{cut}$ & \TA$_\text{cut}$ & \AL$_\text{cut}$\\
      \hline
      VS$_{1}$ & $\leq 1.25$ & $\leq -1.5$ & $\geq 0.45$\\
      VS$_{2}$ & $\leq 1.25$ & $\leq -1.5$ & $\leq 0.45$\\
      VS$_{3}$ & $\geq 1.25$ & $\leq -1.5$ & $\leq 0.45$\\
      VS$_{4}$ & $\leq 1.25$ & $\geq -1.5$ & $\leq 0.45$\\
      VS$_{5}$ & $\geq 1.25$ & $\leq -1.5$ & $\geq 0.45$\\
      VS$_{6}$ & $\leq 1.25$ & $\geq -1.5$ & $\geq 0.45$\\
      VS$_{7}$ & $\geq 1.25$ & $\geq -1.5$ & $\geq 0.45$\\
      Sig S    & $\geq 1.25$ & $\geq -1.5$ & $\leq 0.45$\\
      \hline
    \end{tabular}
    \label{tab:validationregions}
\end{table*}

\section{Results}\label{sec:results}

\begin{figure}[h!]
    \centering
 \includegraphics[width=0.58\linewidth]{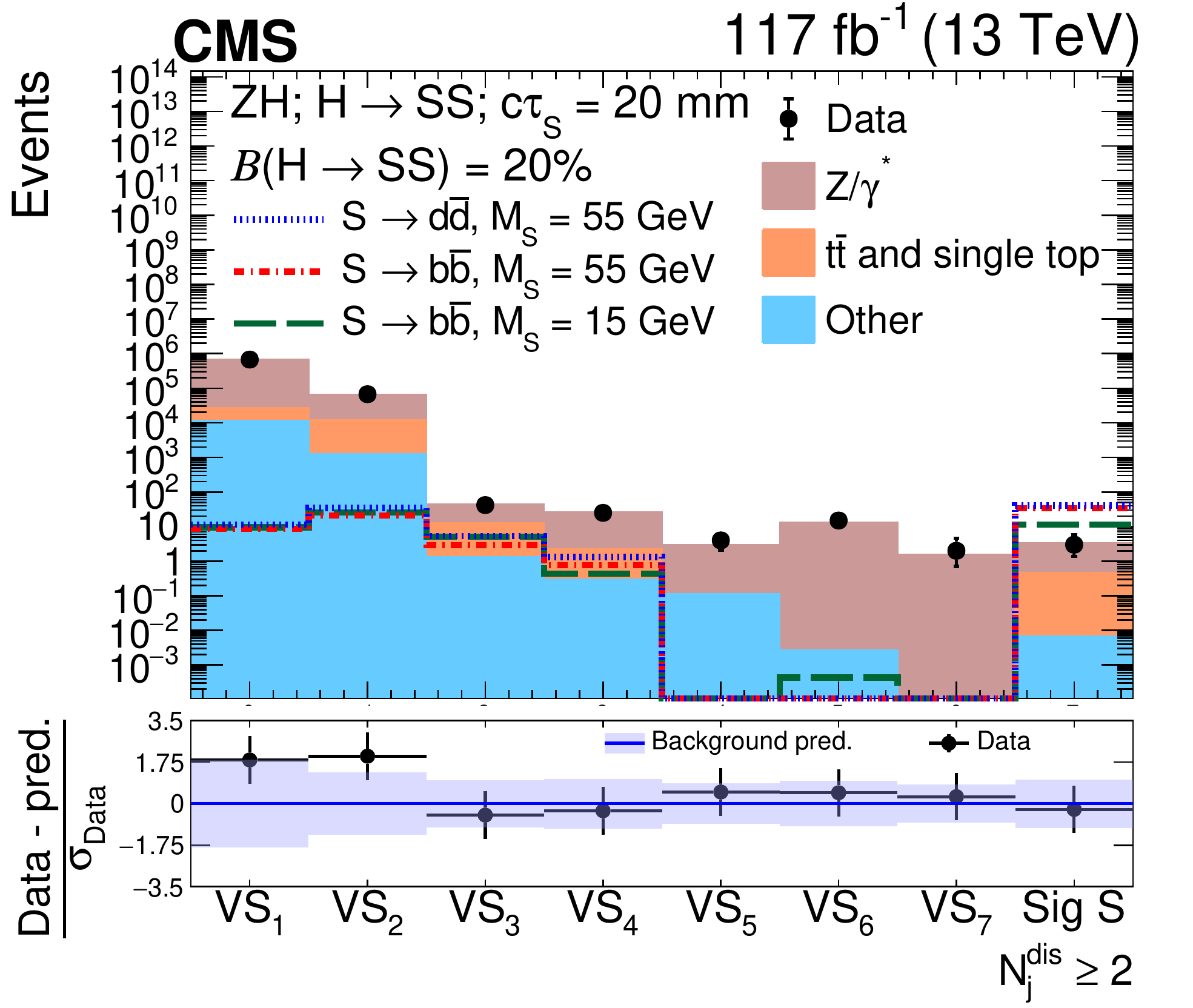}
    \caption{The background estimate and the observed data in the $\NT \geq 2$ bin, for each of the
      seven validation samples (VS$_{1}$ through VS$_{7}$), along with the signal sample (Sig S).
     Signal model distributions for scalar masses of 15 and 55\GeV with a proper mean decay length of 20\mm are also shown.
     The Higgs boson branching fraction to long-lived scalars ($\cbf(\HToSS)$) is set to 20\%.}
    \label{fig:jetmult}
\end{figure}

The result of the background estimation procedure is compared with the observed data in Table~\ref{tab:Ntags} and Fig.~\ref{fig:jetmult}.
The estimated background yields in the three \NT bins, shown in Table~\ref{tab:Ntags}, are obtained using the fit procedure described in Section~\ref{sec:background}.
The content of the $\NT\geq 2$ bin in each of the seven validation samples and the signal sample is shown in Fig.~\ref{fig:jetmult}.
Additional minor background processes, hereafter referred to as ``Other" backgrounds, are included in Fig.~\ref{fig:jetmult}.
``Other" backgrounds include processes such as the production of SM dibosons, SM Higgs bosons, {\PW}+jets, and QCD.
No excess in the data with respect to the SM background is observed.
In the most sensitive bin in the signal sample, $\NT \geq 2$, we observe 3 events with an expected background of $3.5 \pm 1.8$.

\begin{figure}[ht]
    \centering
    \includegraphics[width=0.48\linewidth]{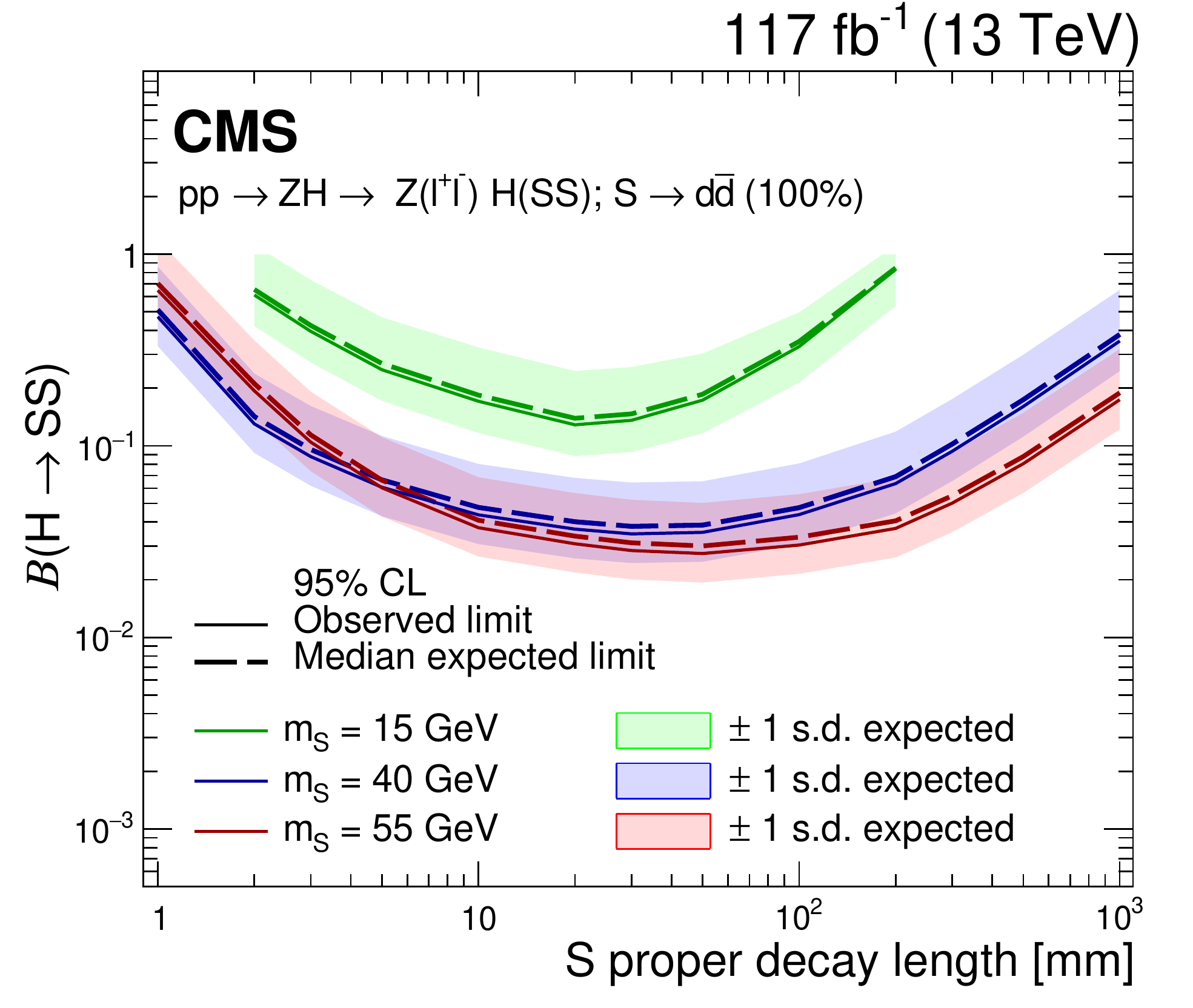}
    \includegraphics[width=0.48\linewidth]{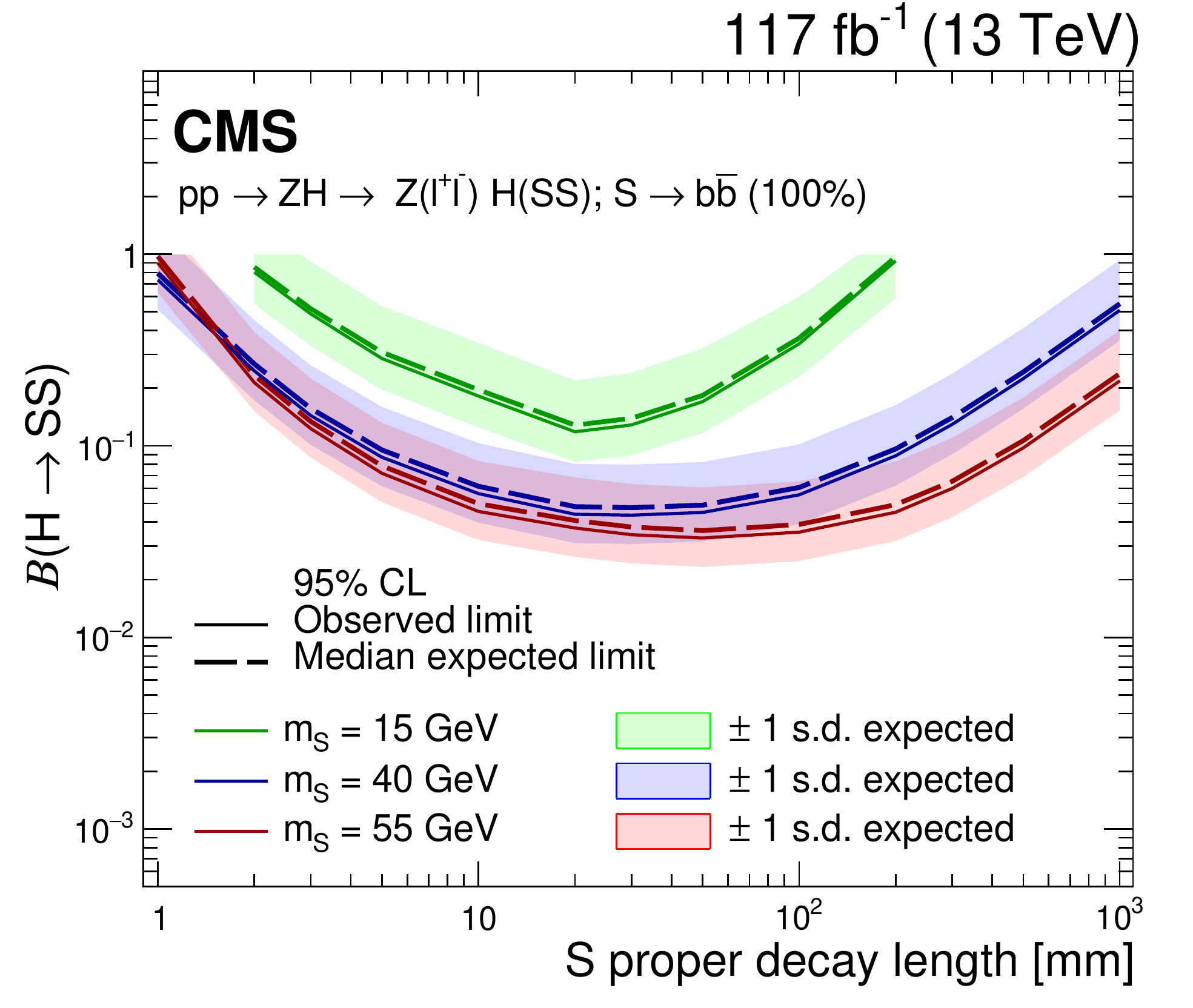}
    \caption{Exclusion limits at 95\% \CL on the Higgs boson branching fraction to long-lived scalars $\cbf(\HToSS)$.
        Limits are presented for scalar decays to \PQd quarks (left) and \PQb quarks (right) as a function of the mean proper decay length of the scalar.
       The limits for the different scalar masses are shown in different colors for each scalar decay mode.}
       \label{fig:limits1}
\end{figure}

\begin{table*}[h!]
\centering
  \topcaption{Observed data and estimated background yields in the search sample for the three \NT bins.
  The estimated background yields are obtained using the fit procedure described in Sec.~\ref{sec:background}.
  Signal yields for two model points are also shown, where the Higgs boson decay branching fraction $\cbf(\HToSS)$ is assumed to be 10\%.}
    \begin{tabular}{l|c|c|c}\hline
      Process & \NT $=$ 0 & \NT $=$ 1  & \NT $\geq$ 2\\
      \hline
      $m_\PS = 15\GeV$, $c\tau_\PS = 20\mm$ & 68.1 $\pm$ 5.1 & 35.6 $\pm$ 5.8 & 5.7 $\pm$ 1.9    \\
      $m_\PS = 55\GeV$, $c\tau_\PS = 20\mm$ & 45.4  $\pm$ 3.4  & 41.0 $\pm$ 6.7 & 16.8 $\pm$ 5.5   \\
      \hline
      Predicted background & $(2\,586.7 \pm 2.6)\times 10^3$ & 4\,035 $\pm$ 67     & 3.5 $\pm$ 1.8   \\
      \hline
      \hline
      Data & 2\,586\,768 & 4\,038 & 3\\
      \hline
    \end{tabular}
    \label{tab:Ntags}
\end{table*}

We set upper limits on the Higgs boson branching fraction to a pair of long-lived scalars for different masses and as a function of the mean proper decay length of the scalar.
We follow the LHC \CLs criterion~\cite{CLS2,CLS1} by using the profile likelihood ratio test statistic and the asymptotic formula~\cite{Asymptotic} to evaluate the 95\% confidence level (\CL) observed and expected limits on the Higgs boson branching fraction to a pair of long-lived scalars.
Systematic uncertainties are propagated by incorporating nuisance parameters that represent the different sources of the uncertainty, which are profiled in the maximum likelihood fit~\cite{CMS-NOTE-2011-005}.
Two scenarios are considered: \PS decays to a pair of \PQb quarks and \PS decays to a pair of \PQd quarks.
The 95\% \CL upper limits are shown in Fig.~\ref{fig:limits1}.
We constrain the Higgs boson branching fraction to long-lived scalars decaying to \PQd (\PQb) quarks at the 3--4 (4--5)\% level for masses of 40 and 55\GeV and mean proper decay lengths in the range 10--100\mm.
The upper bound on the Higgs boson branching fraction to 15\GeV long-lived scalars decaying to \PQd (\PQb) quarks is about 14 (13)\% in the 20--35\mm range of mean proper decay lengths.
In addition to interpreting the results in the context of exotic Higgs boson decays to long-lived scalar particles, we note that this search is sensitive to beyond the SM scenarios in which LLPs are produced in association with a \PZ boson.
The detection efficiency for such LLPs with $c\tau$ of 5--100\mm ranges from 1 to 10\% per LLP depending on its mass.
Heavier LLPs have a larger detection efficiency.

\section{Summary}\label{sec:summary}
A search for long-lived particles (LLPs) produced in association with a \PZ boson decaying to a pair of electrons or muons has been performed.
The decays of LLPs in the tracker volume result in a displaced-jet signature, which is used to distinguish signal from SM background.
No excess over the expected SM event rate is observed.
The results of this search provide sensitivity to beyond the SM scenarios that predict LLPs produced in association with a \PZ boson.
In particular, stringent exclusion limits on exotic Higgs boson decays to long-lived scalars are obtained.
The Higgs boson decay branching fraction $\cbf(\HToSS)$ is constrained to values less than 6\% for proper decay lengths of 10--100\mm and for long-lived particle masses between 40 and 55\GeV.
In the case of low-mass ($\approx$15\GeV) scalar particles that subsequently decay to a pair of \PQb quarks, the search is sensitive to branching fractions $\cbf(\HToSS) < 20\%$ for mean proper decay lengths of 10--50\mm.
This corresponds to an improvement in sensitivity with respect to gluon fusion searches by a factor of 2 at a proper decay length of 30\mm.

\begin{acknowledgments}
  We congratulate our colleagues in the CERN accelerator departments for the excellent performance of the LHC and thank the technical and administrative staffs at CERN and at other CMS institutes for their contributions to the success of the CMS effort. In addition, we gratefully acknowledge the computing centers and personnel of the Worldwide LHC Computing Grid and other centers for delivering so effectively the computing infrastructure essential to our analyses. Finally, we acknowledge the enduring support for the construction and operation of the LHC, the CMS detector, and the supporting computing infrastructure provided by the following funding agencies: BMBWF and FWF (Austria); FNRS and FWO (Belgium); CNPq, CAPES, FAPERJ, FAPERGS, and FAPESP (Brazil); MES and BNSF (Bulgaria); CERN; CAS, MoST, and NSFC (China); MINCIENCIAS (Colombia); MSES and CSF (Croatia); RIF (Cyprus); SENESCYT (Ecuador); MoER, ERC PUT and ERDF (Estonia); Academy of Finland, MEC, and HIP (Finland); CEA and CNRS/IN2P3 (France); BMBF, DFG, and HGF (Germany); GSRI (Greece); NKFIA (Hungary); DAE and DST (India); IPM (Iran); SFI (Ireland); INFN (Italy); MSIP and NRF (Republic of Korea); MES (Latvia); LAS (Lithuania); MOE and UM (Malaysia); BUAP, CINVESTAV, CONACYT, LNS, SEP, and UASLP-FAI (Mexico); MOS (Montenegro); MBIE (New Zealand); PAEC (Pakistan); MSHE and NSC (Poland); FCT (Portugal); JINR (Dubna); MON, RosAtom, RAS, RFBR, and NRC KI (Russia); MESTD (Serbia); SEIDI, CPAN, PCTI, and FEDER (Spain); MOSTR (Sri Lanka); Swiss Funding Agencies (Switzerland); MST (Taipei); ThEPCenter, IPST, STAR, and NSTDA (Thailand); TUBITAK and TAEK (Turkey); NASU (Ukraine); STFC (United Kingdom); DOE and NSF (USA).

  \hyphenation{Rachada-pisek} Individuals have received support from the Marie-Curie program and the European Research Council and Horizon 2020 Grant, contract Nos.\ 675440, 724704, 752730, 758316, 765710, 824093, 884104, and COST Action CA16108 (European Union); the Leventis Foundation; the Alfred P.\ Sloan Foundation; the Alexander von Humboldt Foundation; the Belgian Federal Science Policy Office; the Fonds pour la Formation \`a la Recherche dans l'Industrie et dans l'Agriculture (FRIA-Belgium); the Agentschap voor Innovatie door Wetenschap en Technologie (IWT-Belgium); the F.R.S.-FNRS and FWO (Belgium) under the ``Excellence of Science -- EOS" -- be.h project n.\ 30820817; the Beijing Municipal Science \& Technology Commission, No. Z191100007219010; the Ministry of Education, Youth and Sports (MEYS) of the Czech Republic; the Deutsche Forschungsgemeinschaft (DFG), under Germany's Excellence Strategy -- EXC 2121 ``Quantum Universe" -- 390833306, and under project number 400140256 - GRK2497; the Lend\"ulet (``Momentum") Program and the J\'anos Bolyai Research Scholarship of the Hungarian Academy of Sciences, the New National Excellence Program \'UNKP, the NKFIA research grants 123842, 123959, 124845, 124850, 125105, 128713, 128786, and 129058 (Hungary); the Council of Science and Industrial Research, India; the Latvian Council of Science; the Ministry of Science and Higher Education and the National Science Center, contracts Opus 2014/15/B/ST2/03998 and 2015/19/B/ST2/02861 (Poland); the Funda\c{c}\~ao para a Ci\^encia e a Tecnologia, grant CEECIND/01334/2018 (Portugal); the National Priorities Research Program by Qatar National Research Fund; the Ministry of Science and Higher Education, projects no. 14.W03.31.0026 and no. FSWW-2020-0008, and the Russian Foundation for Basic Research, project No.19-42-703014 (Russia); the Programa Estatal de Fomento de la Investigaci{\'o}n Cient{\'i}fica y T{\'e}cnica de Excelencia Mar\'{\i}a de Maeztu, grant MDM-2015-0509 and the Programa Severo Ochoa del Principado de Asturias; the Stavros Niarchos Foundation (Greece); the Rachadapisek Sompot Fund for Postdoctoral Fellowship, Chulalongkorn University and the Chulalongkorn Academic into Its 2nd Century Project Advancement Project (Thailand); the Kavli Foundation; the Nvidia Corporation; the SuperMicro Corporation; the Welch Foundation, contract C-1845; and the Weston Havens Foundation (USA).
\end{acknowledgments}

\bibliography{auto_generated}
\cleardoublepage \appendix\section{The CMS Collaboration \label{app:collab}}\begin{sloppypar}\hyphenpenalty=5000\widowpenalty=500\clubpenalty=5000\vskip\cmsinstskip
\textbf{Yerevan Physics Institute, Yerevan, Armenia}\\*[0pt]
A.~Tumasyan
\vskip\cmsinstskip
\textbf{Institut f\"{u}r Hochenergiephysik, Vienna, Austria}\\*[0pt]
W.~Adam, J.W.~Andrejkovic, T.~Bergauer, S.~Chatterjee, M.~Dragicevic, A.~Escalante~Del~Valle, R.~Fr\"{u}hwirth\cmsAuthorMark{1}, M.~Jeitler\cmsAuthorMark{1}, N.~Krammer, L.~Lechner, D.~Liko, I.~Mikulec, P.~Paulitsch, F.M.~Pitters, J.~Schieck\cmsAuthorMark{1}, R.~Sch\"{o}fbeck, D.~Schwarz, S.~Templ, W.~Waltenberger, C.-E.~Wulz\cmsAuthorMark{1}
% ...
\vskip\cmsinstskip
% ...
1:  Also at TU Wien, Wien, Austria\\
% ...
 
\end{sloppypar}
\end{document}